# Energy threshold in Smith-Purcell radiation

Sunchao Huang[1,2,3,†], Xihang Shi[4,†], Xiaoqiuyan Zhang[1], Suguo Chen[5], Yue Wang[5], Shengpeng Yang[1,2,3], Ping Zhang[1,2,3], Min Hu[1], and Yubin Gong[1,2,3*]

[1]*School of Electronic Science and Engineering, University of Electronic Science and Technology of China, Chengdu, Sichuan 611731, China*
[2]*Terahertz Radiation and Application Key Laboratory of Sichuan Province*
[3]*National Key Laboratory of Science and Technology on Vacuum Electronics, University of Electronic Science and Technology of China, Chengdu, China*
[4]*Solid State Institute and Faculty of Electrical and Computer Engineering, Technion–Israel Institute of Technology, Haifa 3200003, Israel*
[5]*Department of Applied Physics, Xian University of Technology, Xian 710048, China*
[†]*These authors contributed equally to this work.*
*\*ybgong@uestc.edu.cn*

**Smith-Purcell radiation has emerged as a crucial platform for investigating light-matter interactions and developing compact, tunable light sources that span from microwaves to X-rays. In classical theory, it is believed that Cherenkov radiation exhibits an energy threshold for electrons, while Smith-Purcell radiation is considered free of such a threshold. Although quantum theory suggests there is an emission cutoff in Smith-Purcell radiation, the behavior of this radiation near the threshold remains understudied. In this article, we address this gap by examining the behavior of Smith-Purcell radiation near the threshold from quantum perspectives. Specifically, we derive a quantum energy threshold based on energy-momentum conservation, providing a rigorous limit for the onset of Smith-Purcell radiation. Furthermore, we find that around the threshold the incident electron emits a photon and subsequently reverses its direction of motion. Additionally, we establish a classical energy threshold—below which the classical theory breakdown—by applying the Duane-Hunt limit to Smith-Purcell radiation. Accordingly, quantum theory is required when the electron energy falls between the classical and quantum thresholds. Our findings enrich the understanding of Smith-Purcell radiation and provide valuable insights for developing low-energy-driven and heralded quantum light sources.**

Free electrons emit light spontaneously when passing through or near optical materials. This phenomenon has profound implications in electron microscopy [1,2], light sources [3–6], and high-energy particle detectors [7]. Recent rapid advances in nanophotonics [8] have opened new avenues in this century-old field, particularly with semi-relativistic and low-energy electrons. There are several advantages of using low-energy electrons. First, compact light sources, especially those operating at ultrashort wavelengths, have been demonstrated [9–11], due to the reduced footprint enabled by utilizing low-energy electrons. Second, the quantum nature of low-energy electrons [12,10,13,14], which can be accessed in electron microscopes [15–17], has been harnessed to enhance the cross-section of electron radiation [18,19] and generate quantum light [20–23].

Free electrons cannot emit light in vacuum due to the energy-momentum mismatch between free electrons and free photons. One way to resolve this mismatch is increasing the photon momentum through the use of dielectric media, as exemplified by Cherenkov radiation [24–31]. Another approach is providing additional momentum via optical structures, with Smith-Purcell radiation [32–42] being a well-established example. Smith-Purcell radiation is generated when free electrons interact with periodic structures.

Cherenkov radiation in homogeneous materials is an energy-threshold phenomenon, requiring electron velocity to exceed $c/n$, where $c$ is the vacuum light speed and $n$ is the dielectric constant. In contrast, Smith-Purcell radiation is widely believed to be an energy-threshold-free phenomenon. This property makes Smith-Purcell radiation particularly promising for generating ultrashort-wavelength light from compact electron sources [9,10,43–46]. Additionally, Smith-Purcell radiation could be excited by low-energy electrons, presenting significant potential for on-chip light sources [47–49]. To address this potential, recent advances have significantly reduced the electron energy required for Smith-Purcell radiation, from 300 keV [36] in 1953 to 1.5 keV [39] in 2018, and most recently to 300 eV [49] in 2024.

Generally, the photon energy of Smith-Purcell radiation is significantly lower than the kinetic energy of the incident free electrons. This allows for the assumption that the energy loss of free electrons during radiation can be neglected. This assumption is referred to as the nonrecoil approximation. Under this approximation, the energy of the emitted photon can be calculated as follows, [50,51]

$$E_{\text{p}}^{\text{C}} = \frac{2\pi m\hbar c\beta \cos\theta_{\text{til}}}{d(1-\beta\cos\theta_{\text{obs}})}, \qquad (1)$$

where $m$ is the emission order, $\hbar$ is the reduced Planck constant, $\beta$ is the ratio between the velocity of the incident electron and vacuum light speed, $\theta_\text{til}$ is the tilt angle of the grating i.e., the angle between the grating's normal and the incident electron's moving direction (Fig. 1a), $d$ is the grating length, and $\theta_\text{obs}$ is the observation angle respect to the moving direction of the incident electron. The theory described above is known as the classical theory of Smith-Purcell radiation. Equation (1) indicates that there is no threshold on the energy of electrons in Smith-Purcell radiation; a decrease in electron energy leads to an increase in the emitted photon wavelength. However, this equation is derived from classical electrodynamics, which assumes a constant velocity for electrons and neglects the effects of recoil from light emission. While classical electrodynamics has effectively described most free-electron radiation phenomena to date, Vitaly L. Ginzburg pointed out in 1940 that the nonrecoil assumption fails within the framework of quantum theory, leading to deviations between the energy of emitted photons and the values predicted classically. This phenomenon, known as quantum recoil, has been theoretically predicted [14,52] and recently experimentally demonstrated [10] in Smith-Purcell radiation. Consequently, the validity of Equation (1) when considering electron recoil remains uncertain.

Historically, Smith-Purcell radiation was generated by free electrons interacting with gratings whose grating lengths ranged from tens of nanometers to micrometers or more [36,39,40,50,53]. Due to limitations in grating fabrication accuracy, the minimum grating length has been restricted to approximately tens of nanometer [44,49]. This relatively large grating length ensures that the energy of the emitted photons is much lower than that of the incident electrons, therefore the nonrecoil assumption is justified. Recently, van der Waals structures have demonstrated versatile properties as atomic-scale gratings for Smith-Purcell radiation [9,10,54–57]. The interlayer distance between adjacent van der Waals layers is in the sub-nanometer range, resulting in violations of the nonrecoil assumption. Consequently, quantum recoil has been observed in low-energy-electron-driven van der Waals materials [10].

Here, we revisit the properties of Smith-Purcell radiation within the quantum framework, where the conventionally adopted nonrecoil assumption has been eliminated. By applying energy–momentum conservation, we calculate the emitted photon energy and determine a lower limit of incident electron energy for the onset of Smith-Purcell radiation, referred to as the quantum energy threshold $E_k^Q$. Additionally, by applying the Duane–Hunt limit, we identify a classical energy threshold $E_k^C$ below which classical theory breaks down. Our findings provide new insights for the development of tunable on-chip free electron sources.

**Results**

When a free electron travels near to or through a periodic structure (grating), Smith-Purcell radiation is emitted (see **Fig. 1a**). Under nonrecoil approximation, the electron maintains a constant velocity near the grating, and the output photon energy is determined by Eq. (1). To go beyond the nonrecoil approximation, we apply energy–momentum conservation to Smith–Purcell radiation, which is the foundation of quantum methods, and obtain

$$E_\text{i} = E_\text{f} + \hbar k c, \tag{2}$$

$$\mathbf{p}_\text{i} = \mathbf{p}_\text{f} + \hbar(\mathbf{k} + \mathbf{g}), \tag{3}$$

where $E_\text{i}$ and $E_\text{f}$ are the initial and final energy of the free electron, $\mathbf{k}$ is wavevector of the emitted light and $k = |\mathbf{k}|$, $\mathbf{p}_\text{i}$ and $\mathbf{p}_\text{f}$ are the initial and final momentum of the free electron, $\mathbf{g}$ is the reciprocal lattice vector of the grating, and $g = |\mathbf{g}|$ and $g = 2\pi m/d$ for a one-dimensional grating. Combining Eq. (2) and Eq. (3), we obtain the energy of the emitted photons as

$$E_\text{p}^\text{Q} = \hbar c^2 \frac{\mathbf{p}_\text{i} \cdot \mathbf{g} - \hbar \mathbf{g}^2/2}{E_\text{i} - c\hat{\mathbf{k}} \cdot (\mathbf{p}_\text{i} - \hbar \mathbf{g})}. \tag{4}$$

Similar results have been previously obtained in [10,52]. Although the emission cutoff phenomenon in Smith-Purcell radiation has been identified based on energy-momentum conservation [10,52], explicit expressions for the momentum and energy thresholds for Smith-Purcell radiation have not yet been established. Notably, Eq. (4) is reduced to the classical Smith-Purcell formula Eq. (1) by letting $\hbar \mathbf{g} = 0$, which is justified under the condition that $|\hbar \mathbf{g}| \ll |\mathbf{p}_\text{i}|$. We obtain the momentum threshold from Eq. (4) by applying the condition that photon energy $E_\text{p}^\text{Q}$ is always larger than zero,

$$p_\text{i} \cos \theta_\text{til} > \frac{\hbar g}{2}. \tag{5}$$

The kinetic energy threshold of Smith-Purcell radiation is far less than the rest energy of electrons, which enables us to represent the kinetic energy of electron as

$$E_\text{k} = \frac{p_\text{i}^2}{2 m_e} = \frac{1}{2} m_e \beta^2 c^2, \tag{6}$$

where $m_e$ is the rest mass of an electron. Combining Eq. 6 and Eq. (7), we obtain the quantum kinetic energy threshold of Smith-Purcell radiation as follows:

$$E_\text{k}^\text{Q} = \frac{\pi^2 m^2 \hbar^2}{2 m_e d^2 \cos^2 \theta_\text{til}}, \tag{7}$$

We note that the energy threshold depends on the grating length $d$, the emission order $m$, and the tilt angle $\theta_\text{til}$. **Figure 1b** shows the energies of emitted photons in Smith-Purcell radiation by free electrons interacting with a graphite thin film at various electron energies. The grating length is 0.6708 nm. The quantum results indicate that Smith-Purcell radiation is prohibited when the electron kinetic energy falls below 13 eV, which is the green region in **Fig. 1b**.

In contrast, the classical results (the red dashed line) do not inherently show such a lower electron energy bound. However, according to the Duane-Hunt limit [58], the maximum energy of the emitted photon equals the kinetic energy of the incident electron, i.e., $E_\text{p}^\text{C} = E_\text{k}^\text{C}$. Furthermore, since the classical kinetic energy threshold $E_\text{k}^\text{C}$ is small in magnitude, we can approximate the kinetic energy using the non-relativistic form. At the Duane–Hunt limit, the electron kinetic energy equals the emitted photon energy, as given by Eq. (1) for Smith–Purcell radiation,

$$\frac{1}{2} m_e \beta^2 c^2 = \frac{2\pi m \hbar c \beta \cos\theta_{\text{til}}}{d(1 - \beta \cos\theta_{\text{obs}})}. \tag{8}$$

We could then obtain $\beta$ from Eq. (8), and subsequently substituting the resulting expression for $\beta$ into the kinetic energy formula, yield

$$E_k^C = \frac{m_e}{8 \cos\theta_{\text{obs}}^2} \left( c - \sqrt{c^2 - 4b \cos\theta_{\text{obs}}} \right)^2, \tag{9}$$

where $b = 4\pi mc\hbar\cos\theta_{\text{til}}/(dm_e)$. Here, we have leveraged the small magnitude of $E_k^C$, which justifies the selection of the negative root for $\beta$. Below this energy threshold, the classical theory predicts emitted photon energies that exceed the kinetic energy of the incident electron, violating the Duane-Hunt limit. Using Eq. (9), we calculate the classical kinetic energy threshold and plot a black vertical dashed line, denoted as $E_k^C$ in **Fig. 1b**. Our findings reveal that there is indeed a lower bound on the electron energy in Smith-Purcell radiation from both classical and quantum perspectives. The classical theory ceases to be valid when the electron energy falls below the former, while no Smith–Purcell radiation occurs when the electron energy is below the latter.

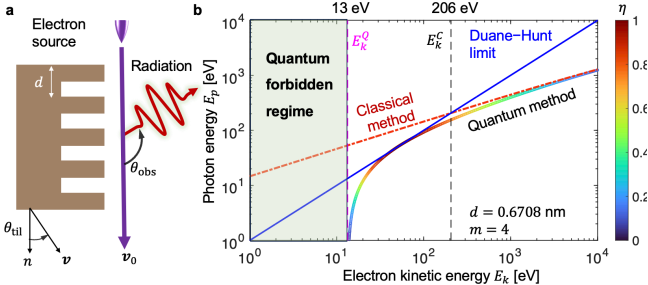

**Fig. 1. Demonstration of a lower bound on the electron energy in Smith-Purcell radiation. (a),** the schematic diagram of Smith-Purcell radiation where electrons move near or pass through a periodic structure (grating) resulting in free electron radiation. The output photon energies of the radiation can be manipulated via changing the grating distance $d$, electron energy and the observation angle $\theta_{\text{obs}}$, providing a promising way to obtain tunable light sources. $\theta_{\text{til}}$ is the angle between the grating's normal vector $\mathbf{n}$ and the electron velocity $\mathbf{v}$. We position the velocity vector intentionally not parallel to the normal vector to illustrate $\theta_{\text{til}}$. **(b)** Demonstration of the quantum energy threshold (rainbow color curve) and the classical energy threshold (red curve) in Smith-Purcell radiation, where the grating length is $d = 0.6708$ nm (the layer distance of graphite), the emission order is 4 and the observation angle $\theta_{\text{obs}} = 130°$ and $\theta_{\text{til}} = 0°$.

In **Fig. 1b**, the quantum energy threshold $E_k^Q$ does not align with the classical threshold $E_k^C$, primarily because energy conservation is not upheld in the classical approach. It is important to note that the quantum method remains valid across the entire range of electron kinetic energies, as it explicitly incorporates energy-momentum conservation. In contrast, the classical method relies on the nonrecoil approximation, assuming that incident electrons do not lose energy during Smith-Purcell radiation. Consequently, the classical method can only accurately predict emitted photon energies when quantum recoil effects are negligible. As shown in **Fig. 1b**, there is a close agreement between the classical results (red curve) and quantum results at electron kinetic energies greater than $8 \times 10^3$ eV, suggesting minimal quantum recoil effects. However, when the electron kinetic energy falls below $8 \times 10^3$ eV, the classical results significantly diverge from the quantum results, indicating substantial quantum recoil effects.

In **Fig. 1b**, while the quantum curve consistently remains below the Duane-Hunt limit, the classical curve exceeds the Duane-Hunt limit when the electron kinetic energy is below the classical threshold $E_k^C$. This phenomenon further confirms that the classical method fails to accurately calculate the photon energies of Smith-Purcell radiation driven by low-energy electrons. As the quantum curve approaches the Duane-Hunt limit, the electrons convert all their kinetic energy into photons via Smith-Purcell radiation. To characterize the electron energy efficiency in Smith-Purcell radiation, we introduce an efficiency parameter $\eta = E_p^Q/E_k$, where $E_k$ is the kinetic energy of the incident electrons. The electron energy efficiency is illustrated in **Fig. 1b** using color coding for the quantum results. In conventional Smith-Purcell radiation, the value of $\eta$ is typically on the order of 0.01% [50], with the highest experimentally achieved value being 0.74% [49]. By using low-energy electrons to interact with van der Waals materials, we have demonstrated that the value of $\eta$ can reach approximately 11% [10].

In the classical description of Smith–Purcell radiation, the electron velocity is assumed unchanged because the radiation is treated as a continuous electromagnetic wave. The energy emitted by a single electron is marginal so recoil can be neglected. By contrast, the Duane–Hunt limit considers X-rays as discrete photons: the maximum photon energy is set by converting the entire electron kinetic energy into a single photon. This limit is not specific to Smith–Purcell radiation but applies to all free-electron radiation effects. When the classically predicted photon energy exceeds this limit, which is the region below the classical threshold $E_k^C$ (see Fig. 1b and **Fig. 2**), the classical method breaks down.

To accurately describe the behavior of Smith-Purcell radiation below the classical threshold, we must exploit the quantum picture. As shown in Fig. 1a, electrons with kinetic energy below the classical threshold can still generate radiation. As we reduce the kinetic energy of the incident electrons, we observe a scenario where the incoming electron converts all its kinetic energy (point B in **Fig. 2a**) and momentum (point B' in **Fig. 2b**) into Smith-Purcell radiation, as depicted in Fig. 2c. This is also the point that where the quantum predicted curve touch the Duane–Hunt limit line in **Fig. 1b**.

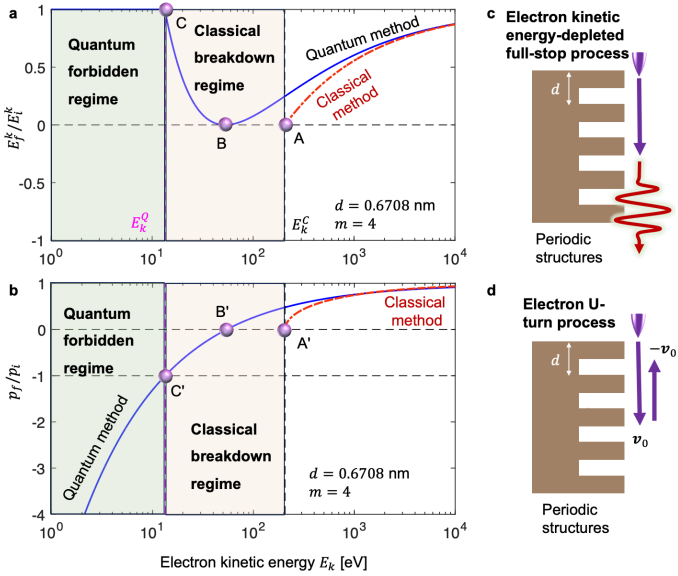

**Fig. 2. Comparison of the Classical and Quantum Threshold Effects of the Smith-Purcell Effect.** This figure presents a comparison of the kinetic energy (a) and momentum (b) of the incident electron after emitting a photon through Smith-Purcell radiation, analyzed through both classical and quantum theories. (c) Illustrates a scenario where the incident electron depletes its entire kinetic energy and comes to a stop after emitting a Smith-Purcell photon. (d) Depicts a scenario at the quantum threshold point, where the incident electron undergoes a U-turn process without losing energy.

When we further reduce the incident electron energy beyond point B, the emitted photon energy is reduced faster than the electron energy, which could be seen from the green curve in **Fig. 1b**. Consequently, the ratio of final electron energy to initial electron energy increases, as shown by curve CB in **Fig. 2a**. What is more interesting is in the momentum space. The final momentum of the incident electron becomes negative compared to its initial momentum in the regime represented by curve C'B', indicating that the incoming electron undergoes a U-turn process.

As the electron energy approaches the quantum threshold $E_k^Q$, the photon energy predicted by quantum theory decreases rapidly (rainbow color curve in **Fig. 1b**), so the energy recoil becomes negligible. However, the momentum recoil cannot vanish, since it is bound below by $\hbar g$. At point C, the electron is elastically scattered backward by the grating without energy loss, as depicted in **Fig. 2d**. The grating then is excited by a static electromagnetic field whose spatial Fourier expansion includes the reciprocal-lattice vector **g**. This elastic scattering is well known in low-energy electron diffraction (LEED), where backscattered electrons are used to probe surface crystal structures.

When the electron energy is below the quantum threshold $E_k^Q$, the electron may still interact with the medium, but not through the Smith–Purcell mechanism. In this regime, processes such as Cherenkov or transition radiation may exist, depending on the material properties. These are material-property-driven radiation phenomena, in contrast to the structure-driven Smith–Purcell radiation.

Due to constraints in grating fabrication accuracy, Smith-Purcell radiation is typically studied using micrometer-scale gratings. The energies of the incident electrons are usually on the order of tens of kilovolts. Although, recent experiments have demonstrated that this value can be reduced to 300 eV [49]. The lowest electron energy used to generate Smith-Purcell radiation is significantly higher than the corresponding quantum energy threshold, as shown in the upper right part of **Fig. 3**, resulting in the long classical belief that there is no lower bound on electron energy in Smith-Purcell radiation and explain the absence of experimental results that challenge this belief. Recently, van der Waals materials have been introduced to generate atomic version Smith-Purcell radiation, which reduces the grating length to the atomic scale. However, the lowest electron energy used to generate this atomic version of Smith-Purcell radiation is also considerably higher than the corresponding quantum energy threshold, as indicated in the upper left part of **Fig. 3**.

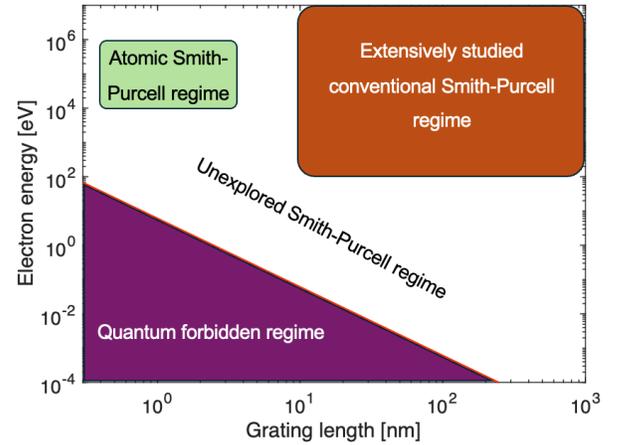

**Fig. 3.** An overview of the different regimes of Smith-Purcell radiation, including conventional Smith-Purcell radiation, atomic version Smith-Purcell radiation, and the quantum forbidden regime.

Equation (1) demonstrates that the classical output photon energy of Smith-Purcell radiation decreases with the increase of the observation angle, akin to the Doppler effect. This behaviour also exists in the classical energy threshold of Smith-Purcell radiation (illustrated by the red curve in **Fig. 4**). This implies that as the kinetic energy of the incident electrons is reduced, the forward Smith-Purcell radiation ($\theta_{obs} < 90°$) becomes prohibited sooner than the backward Smith-Purcell radiation ($\theta_{obs} > 90°$). Interestingly, the quantum energy threshold for Smith-Purcell radiation remains constant across the entire range of observation angles from 0° to 180° (as shown by the green curve in **Fig. 4**). This indicates that Smith-Purcell radiation is simultaneously forbidden in all emission directions if the kinetic energy of the incident electrons falls below a specific threshold, namely, the quantum energy threshold. The physical origin of the differing $\theta_{obs}$ dependencies between the classical and quantum thresholds for Smith-Purcell radiation lies in the quantum recoil effect. Specifically, when the electron energy approaches the threshold values, the quantum recoil effect becomes significant. In contrast, the derivation of the classical energy threshold explicitly neglects quantum recoil effects, which accounts for the discrepancy in their respective behaviors with respect to $\theta_{\text{obs}}$.

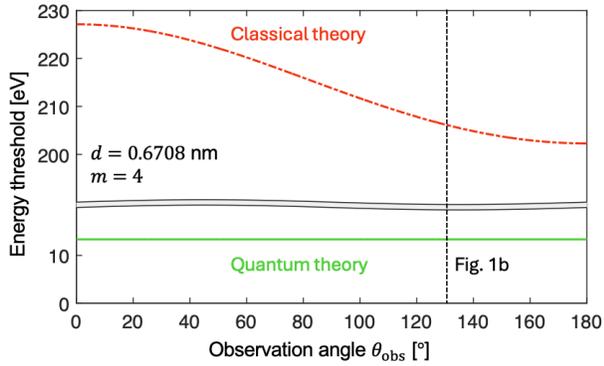

**Fig. 4.** The dependence of classical and quantum energy threshold on the observation angle $\theta_{\text{obs}}$, where the grating length $d = 0.6708$ nm and the emission order $m = 4$. The black dashed line correspondences the observation angle in Fig. 1b.

**Discussion**

Van der Waals materials and heterostructures, including graphite, hexagonal boron nitride (h-BN), molybdenum disulfide ($MoS_2$), and graphite/$MoS_2$, have emerged as versatile platforms for studying Smith-Purcell radiation [9,10,54–57]. This is due to their high in-plane thermal conductivities [59], surfaces free of dangling bonds, and a diverse range of materials with varying interlayer distances [60–62]. The atomic layers of van der Waals materials naturally form gratings, with grating lengths (interlayer distances) ranging from sub-nanometer to a few nanometers, enabling the exploration of Smith-Purcell radiation at sub-nanometer scales. Such sub-nanometer gratings pave the way for compact, tunable X-ray sources generated via Smith-Purcell radiation. Moreover, they offer the potential for experimentally demonstrating quantum recoil in Smith-Purcell radiation [10], a phenomenon proposed by Vitaly L. Ginzburg in 1940.

In this work, we demonstrate that sub-nanometer gratings provide a promising platform for experimentally verifying the energy threshold in Smith-Purcell radiation. As depicted in **Fig. 1** and **Fig. 2**, the classical energy threshold for fourth-order Smith-Purcell radiation at an observation angle of $\theta_{\text{obs}} = 130°$ is 206 eV with a grating length of 0.6708 nm. This threshold can be further increased by reducing the grating length. Electrons with such kinetic energies are routinely available in scanning electron microscopes and have been used to generate Smith-Purcell radiation with 10 nm gratings. Additionally, a 0.6708 nm grating can be achieved using graphite, making the experimental demonstration of the energy threshold in Smith-Purcell radiation feasible in the near future.

In the quantum description of free-electron radiation—including Smith–Purcell radiation—the free electron becomes entangled with the photon it emits [63,64]. This entanglement is enforced by energy–momentum conservation, as expressed in Eqs. (2) and (3). Consequently, the radiation process can be understood as the scattering of a joint electron–photon state. In previous studies, the quantum state of the emitted radiation has been heralded by measuring the final electron energy [65–67]. Here we show that, for low-energy incident electrons, the photon state can be heralded by a simpler approach: detecting the backscattered electron, corresponding to the region BC (B'C') in **Fig. 2**. This reduces experimental complexity and facilitates the integration of an electron-heralded quantum light source, since only the electron's propagation direction needs to be measured rather than its small energy shift.

In summary, we explicitly derive both the quantum and classical energy thresholds for Smith–Purcell radiation. We then investigate the behavior of the radiation near these thresholds from both perspectives. Our analysis shows that the classical theory breaks down when the electron energy falls below the classical threshold, while Smith–Purcell radiation is entirely forbidden when the electron energy lies below the quantum threshold. We also highlight a scenario in which an incident electron emits a photon via Smith-Purcell radiation and subsequently reverses its direction of motion. Notably, at the quantum energy threshold, no Smith-Purcell radiation occurs while the incident electron undergoes a U-turn process without energy loss. Our findings offer valuable insights for developing low-energy-driven and heralded quantum light sources.

**Funding.** This work is supported by the National Natural Science Foundation of China (Grant No. 92463308, 92163204 and 62427806).

**Disclosures.** The authors declare no conflicts of interest.

**Data Availability.** The data that support the findings of this study are available from the corresponding author upon reasonable request.